\documentclass[crystals,article,accept,moreauthors,pdftex]{Definitions/mdpi} 

\firstpage{1} 
\pubvolume{1}
\issuenum{1}
\articlenumber{0}
\pubyear{2021}
\copyrightyear{2021}
\datereceived{} 
\dateaccepted{} 
\datepublished{} 
\hreflink{https://doi.org/} 

\usepackage{graphicx}
\Title{Absence of superconductivity in the Hubbard dimer model for $\kappa$-(BEDT-TTF)$_2$X}
\TitleCitation{Title}

\Author{D. Roy $^{1}$, R. T. Clay $^{1,*}$, S. Mazumdar $^{2}$ }

\AuthorNames{D. Roy, R. T. Clay, S. Mazumdar }

\AuthorCitation{Roy, D.; Clay, R. T.; Mazumdar, S.}

\address{%
  $^{1}$ \quad Department of Physics \& Astronomy and
  HPC$^2$ Center for Computational Sciences, Mississippi State University,
  Mississippi State, MS 39762; r.t.clay@msstate.edu\\
  $^{2}$ \quad Department of Physics, University of Arizona,
  Tucson, AZ 85721; sumit@physics.arizona.edu}

\corres{Correspondence: r.t.clay@msstate.edu} 

\abstract{In the most studied family of organic superconductors
  $\kappa$-(BEDT-TTF)$_2$X, the BEDT-TTF molecules that make up the
  conducting planes are coupled as dimers. For
  some anions X, an antiferromagnetic insulator is found at low
  temperatures adjacent to superconductivity. With an average of one
  hole carrier per dimer, the BEDT-TTF band is effectively
  $\frac{1}{2}$-filled.  Numerous theories have suggested that
  fluctuations of the magnetic order can drive superconducting pairing
  in these models, even as direct calculations of superconducting
  pairing in monomer $\frac{1}{2}$-filled band models find no superconductivity.
  Here we present accurate zero-temperature Density Matrix
  Renormalization Group (DMRG) calculations of a dimerized lattice
  with one hole per dimer. While we do find an antiferromagnetic state
  in our results, we find no evidence for superconducting pairing.
  This further demonstrates that magnetic fluctuations in the
  effective $\frac{1}{2}$-filled band approach do not drive
  superconductivity in these and related materials.}

\begin{document}

\section{Introduction}

Correlated-electron superconductivity (SC) continues to be one of the
most challenging problems in condensed matter physics.  Even as the
large majority of investigators in this research area have focused on
cuprates and iron-based superconductors (and more recently,
ruthenates), it is now widely recognized that there exist many other
unconventional or correlated-electron superconductors, although with
lower superconducting critical temperature T$_c$
\cite{Mazumdar89a,Uemura91a,Fukuyama87a,Fukuyama90a,McKenzie97a,Iwasa03a,Capone09a,Mazumdar12a,Baskaran16a}.
Superconducting organic charge-transfer (CT) solids hold a special
place in this context, as SC in CT solids was discovered significantly
prior to the discovery of the phenomenon in the cuprates
\cite{Jerome80a}.  In spite of intensive investigations over the past
four decades, however, the mechanism of SC in CT solids remains a
mystery, and organic SC continues to be an active research area, as
illustrated by the appearance of many review articles on the subject
over the past decade
\cite{Kanoda11a,Kato12a,Brown15a,Clay19a,Dressel20a}. In the present
paper we report the results of our latest calculations that have
probed a key question in the field of organic SC.

The common theme among many (though not all, see below)
correlated-electron superconductors is the proximity of SC to
spin-density wave (SDW) or antiferromagnetism (AFM) . This property is
shared by organic superconductors based on the cation TMTSF as well as
some $\kappa$-phase BEDT-TTF compounds. Unlike in the cuprates,
though, transition from the insulating magnetic phase to the
superconducting phase is driven by the application of moderate
pressure as opposed to doping. Both TMTSF and BEDT-TTF based
superconductors (hereafter (TMTSF)$_2$X and (BEDT-TTF)$_2$X) are
characterized by cation:anion ratio of 2:1, with closed-shell
monovalent anions X$^-$, the implication being that each individual
cation monomer has charge +$\frac{1}{2}$. In both families, the cation
lattice is highly dimerized, such that the overall charge on a dimer
of cations is exactly +1. This has led to various {\it effective}
half-filled band theories of organic CT superconductors, within which
the dimer of cations is considered as a single site in a lattice
described within anisotropic triangular lattice Hubbard models (we
emphasize that these theories are therefore not applicable to {\it
  nondimerized} (BEDT-TTF)$_2$X families, such as those belonging to
the $\theta$-family, which also exhibit SC; we revisit this issue
later).  Within mean-field and dynamic mean field theory (DMFT)
calculations on the half-filled band anisotropic triangular lattice
Hubbard model, pressure increases the lattice frustration, causing
thereby a bandwidth-driven transition from the AFM insulator to the
superconducting state (or in the case of a lattice that is already
strongly frustrated, as in $\kappa$-(BEDT-TTF)$_2$Cu$_2$(CN)$_3$, from
a quantum spin liquid (QSL) to SC)
\cite{Kino98a,Schmalian98a,Kondo98a,Vojta99a,Baskaran03a,Liu05a,Kyung06a,Yokoyama06a,Watanabe06a,Sahebsara06a,Nevidomskyy08a,Sentef11a,Hebert15a}.
Direct numerical calculations of superconducting pair-pair
correlations within the $\frac{1}{2}$-filled Hubbard band on
triangular lattices have, however, consistently found absence of SC
\cite{Clay08a,Dayal12a,Watanabe08a,Tocchio09a,Gomes13a}. Taken
together with the very recent convincing demonstration of the absence
of SC within weakly doped the square lattice two-dimensional (2D)
Hubbard Hamiltonian \cite{Qin20a}, numerical calculations then cast
serious doubt on the existing approximate spin fluctuation mechanisms
of AFM-to-SC transition.

Conceivably, one possible origin of disagreement between the
mean-field theories of organic SC and numerical calculations is that
retention of the explicit dimeric structure of the Hubbard lattice
sites is essential for superconducting correlations to dominate.  That
is, even as intra-dimer charge fluctuations are precluded within the
effective $\frac{1}{2}$-filled band models (since that would make the
effective $\frac{1}{2}$-filled description invalid and can even
destroy the AFM in the weakly frustrated region of the Hamiltonian),
 the superconducting singlets should be defined such that
they span over a {\it pair of dimers}, as opposed to a pair of
effective single sites. Needless to say such a numerical calculation
is far more involved than where dimers are replaced with single sites,
as the required lattice sizes needed are twice that in the previous
calculations \cite{Clay08a,Dayal12a,Gomes13a}.  We report precisely
such a calculation here, which again finds absence of SC in such a
dimer lattice.

\section{Methods}

\begin{figure}
  \centerline{\includegraphics[width=2.5in]{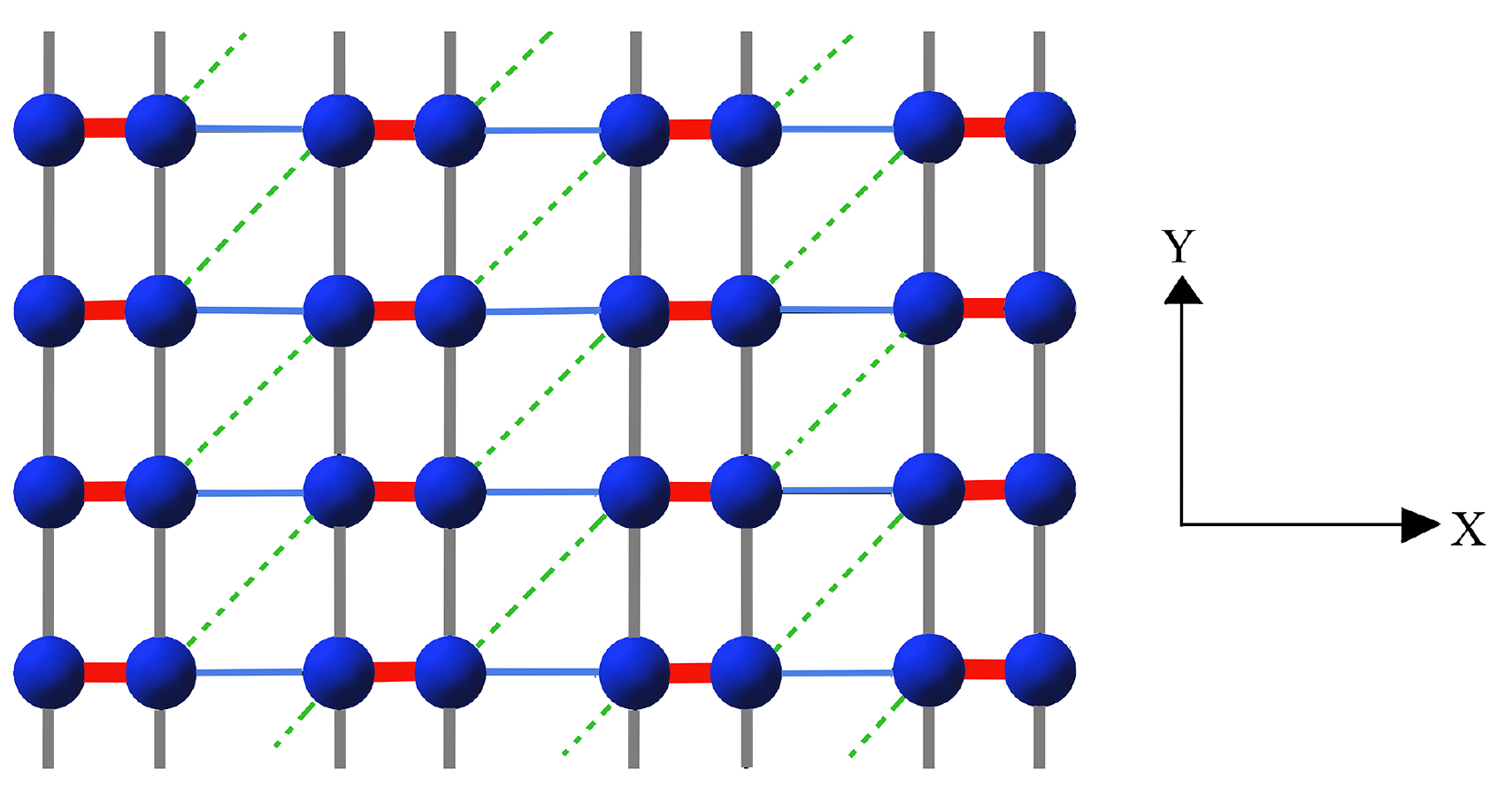}}
  \caption{The dimer lattice we consider. Thick lines are the intra-dimer
    hopping $t_d$, thin lines are the inter-dimer hopping $t$, and
    dashed lines are the frustrating bond $t^\prime$. Boundary conditions
    in our DMRG calculations
  are open in $x$ and periodic in $y$.}
\label{lattice}  
\end{figure}

We consider the  Hubbard Hamiltonian,
\begin{equation}
  H = - \sum_{\langle i,j\rangle,\sigma}  t_{ij} (t^\dagger_{i,\sigma}t_{j,\sigma}+H.c.) + U\sum_i n_{i,\uparrow}n_{i,\downarrow}.
  \label{ham}
\end{equation}
In Eq.~\ref{ham} $c^\dagger_{i,\sigma}$ creates an electron of spin
$\sigma$ on molecule $i$ and
$n_{i,\sigma}=c^\dagger_{i,\sigma}c_{i,\sigma}$. $U$ is the onsite
Hubbard interaction. It is important to note that this $U$ is the
Coulomb repulsion between carriers on each {\it monomer} site and is
not the same as the effective dimer $U_d$.  The lattice we consider is
a rectangular lattice of dimers (see Fig.~\ref{lattice}).  In this
lattice there are three hopping integrals $t_{ij}$: $t_d$ is the
intra-dimer hopping, $t$ the inter-dimer hopping, and $t^\prime$ a
frustrating hopping (see Fig.~\ref{lattice}). We take the system to be
$\frac{1}{4}$ filled with an average electron density of $\frac{1}{2}$
per monomer site. This lattice is expected to show N\'eel AFM order in
the unfrustrated limit of small $t^\prime/t$.

To solve for the ground state of Eq.~\ref{ham} we employ the Density
Matrix Renormalization Group (DMRG) method \cite{White92a} using the
ITensor library \cite{itensor} with real-space parallelization
\cite{Stoudenmire13a}. DMRG can calculate essentially numerically
exact correlation functions in quasi-one-dimensional systems. However,
its primary limitation is an exponential scaling in the transverse
size of systems that can be studied. Here we consider lattices of
dimension 4 sites in $y$ and up to 16 sites in $x$ (see Fig.~\ref{lattice}),
corresponding to an 8$\times$4 lattice in terms of dimers.
We choose $t_d$=1.5 and $t$=0.5 such that the average hopping along
the $x$ axis is 1.0; our choice of $t_d\sim 3 t$ is comparable to
the dimerization found in the CT solid superconductors \cite{Clay19a}. We take
boundary conditions that are open (periodic) in $x$ ($y$). We used
a DMRG bond dimension $m$ of up to 15,000 with minimum truncation
error of 10$^{-8}$-10$^{-7}$; all results we show are extrapolated to
zero truncation error.

\section{Results}

\begin{figure}
  \centerline{\includegraphics[width=3.5in]{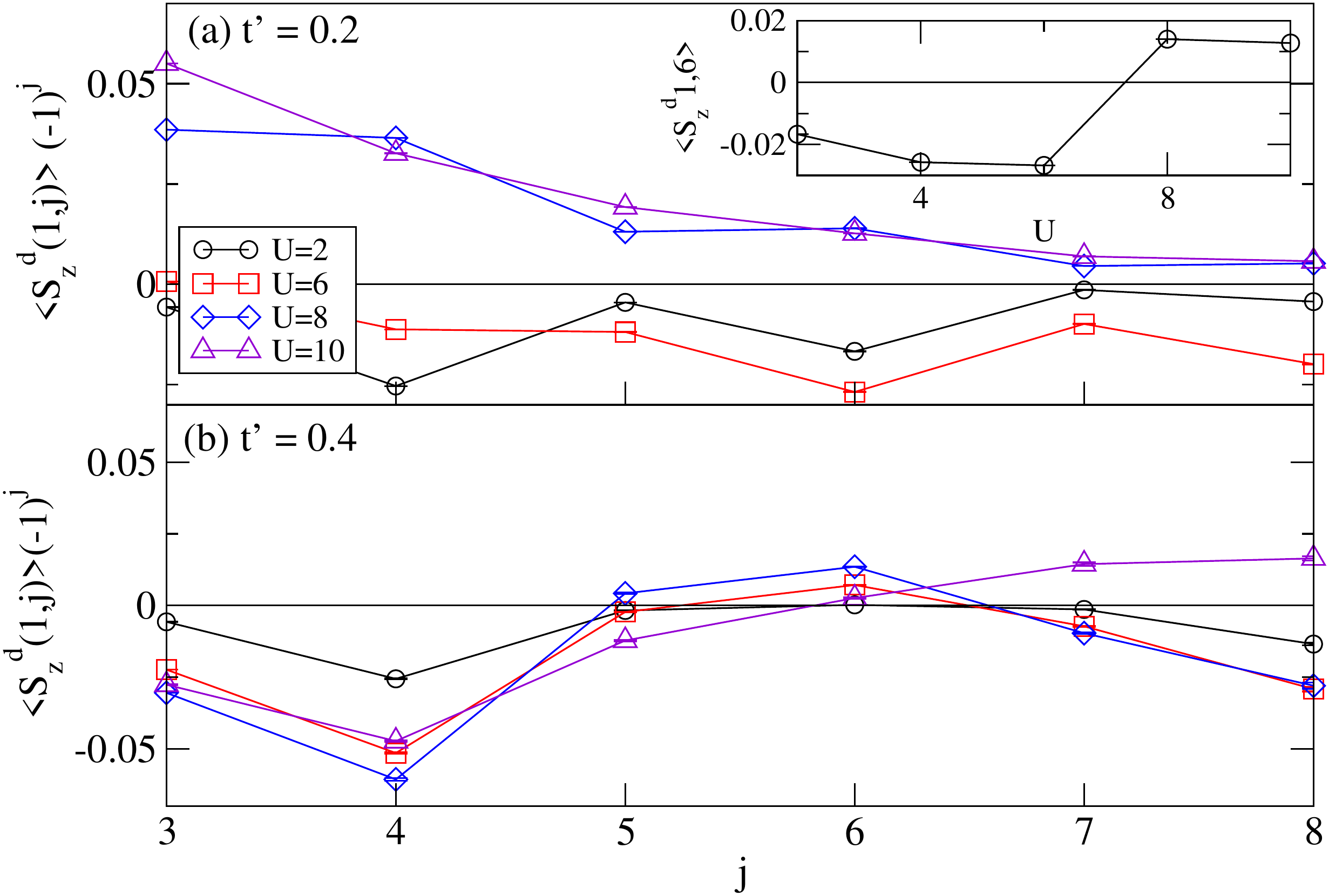}}
  \caption{Dimer spin-spin correlations $\langle S_z^d(1,j)\rangle$
    for a 16$\times$4 lattice between the first dimer on chain 1 and
    dimer $j$ on chain 2, multiplied by $(-1)^j$, the expected sign
    for N\'eel AFM.  (a) $t^\prime=0.2$. The inset shows the $U$
    dependence of a single spin-spin correlation at long distance.
    N\'eel AFM order is present at $t^\prime$ for $U\gtrapprox 8$.
    (b) $t^\prime=0.4$. Here we find no AFM order.}
  \label{spin-spin}
\end{figure}

We measure dimer spin-spin and pair-pair correlation functions.
The dimer spin operator is
$$
n^d_{i,\sigma}=n_{i_1,\sigma}+n_{i_2,\sigma},
$$ where the sites $i_1$ and $i_2$ make up the dimer $i$. We calculate
the dimer spin-spin correlation function $S^d_z(i,j)=\langle
(n^d_{i,\uparrow}-n^d_{i,\downarrow})(n^d_{j,\uparrow}-n^d_{j,\downarrow})\rangle$.
In Fig.~\ref{spin-spin} we plot $\langle S_z^d(1,j)\rangle (-1)^j$
where dimer 1 is on the first chain and dimer $j$ is on the
neighboring chain for $t^\prime=0.2$ and 0.4.
As shown in Fig.~\ref{spin-spin}(a), for $t^\prime=0.2$ and
$U\gtrapprox 8$ this quantity is positive for all spin correlations,
indicating the presence of N\'eel AFM order.  For $t^\prime=0.4$ we do
not find N\'eel AFM order up to at least $U=12$. These results show
that N\'eel AFM order is only present in this system when the
frustration is low, and for realistic $U$ values AFM will not be
present when $t^\prime/t \gtrapprox 0.4$.

To define
superconducting pair-pair correlations in the effective $\frac{1}{2}$-filled
representation, we first define an operator creating an effective  particle
on dimer $i$ with equal monomer populations, as required within
effective $\frac{1}{2}$-filled band theories,
\begin{equation}
d^\dagger_{i,\sigma}=\frac{1}{\sqrt{2}}\left(c^\dagger_{i_1,\sigma}+c^\dagger_{i_2,\sigma}\right).
\end{equation}
The $\Delta^\dagger_{i,i+\delta}$ operator creates a singlet pair
between nearest-neighbor (n.n.) dimers separated by $\delta$:
\begin{equation}
\Delta^\dagger_{i,i+\delta}=\frac{1}{\sqrt{2}}\left(d^\dagger_{i,\uparrow}d^\dagger_{i+\delta,\downarrow}
-d^\dagger_{i,\downarrow}d^\dagger_{i+\delta,\uparrow}\right).
\label{delta}
\end{equation}
Note that Eq.~\ref{delta} involves {\it four} different lattice sites.
We consider two kinds of n.n. pairs with $\delta$ taken as the
n.n. distance between dimers in the $x$ and $y$ directions,
\begin{equation}
  P(r=|\vec{r_i}-\vec{r_j}|)_{\delta,\delta^\prime}=\langle \Delta^\dagger_{i,i+\delta}\Delta_{j,j+\delta^\prime}\rangle.
\label{pr}
\end{equation}
In the effective dimer model pairing is expected to have $d$-wave
symmetry. Because of the small transverse dimensions of our lattice we
do not define full $d_{x^2-y^2}$ pairs involving four n.n. singlets, but
instead calculate only single singlet-singlet correlations. To check
if the pairing has the expected $d$-wave sign structure we calculate
two different correlation functions, $P_{||}(r)\equiv P(r)_{\delta=x,\delta^\prime=x}$
and $P_\perp(r)\equiv P(r)_{\delta=x,\delta^\prime=y}$.
$P_{||}(r)$ corresponds to correlations between two singlets
both oriented along $x$, while $P_\perp(r)$ corresponds to one singlet oriented
along $x$ and one along $y$. For $d_{x^2-y^2}$ pairing
$P_\perp(r)$ should be negative.

In a 2D superconductor long-range order should be present in
$P(r)$ at zero temperature. Because of the quasi-one-dimensional
nature of cylindrical geometry of
the lattice that we can solve in DMRG long-range order in $P(r)$
is not possible even in the limit $x\rightarrow\infty$. If such
a quasi-one-dimensional system showed a tendency to SC however,
the system would behave as a Luther-Emery liquid \cite{Luther74a}
with $P(r)$ decaying as a power law $r^{-\alpha}$ with $\alpha<1$.
In addition to long-range order in $P(r)$, if SC in the system
is driven by Coulomb interactions, one expects that $P(r,U)>P(r,U=0)$
\cite{Clay08a}.

\begin{figure}
  \centerline{\includegraphics[width=4.75in]{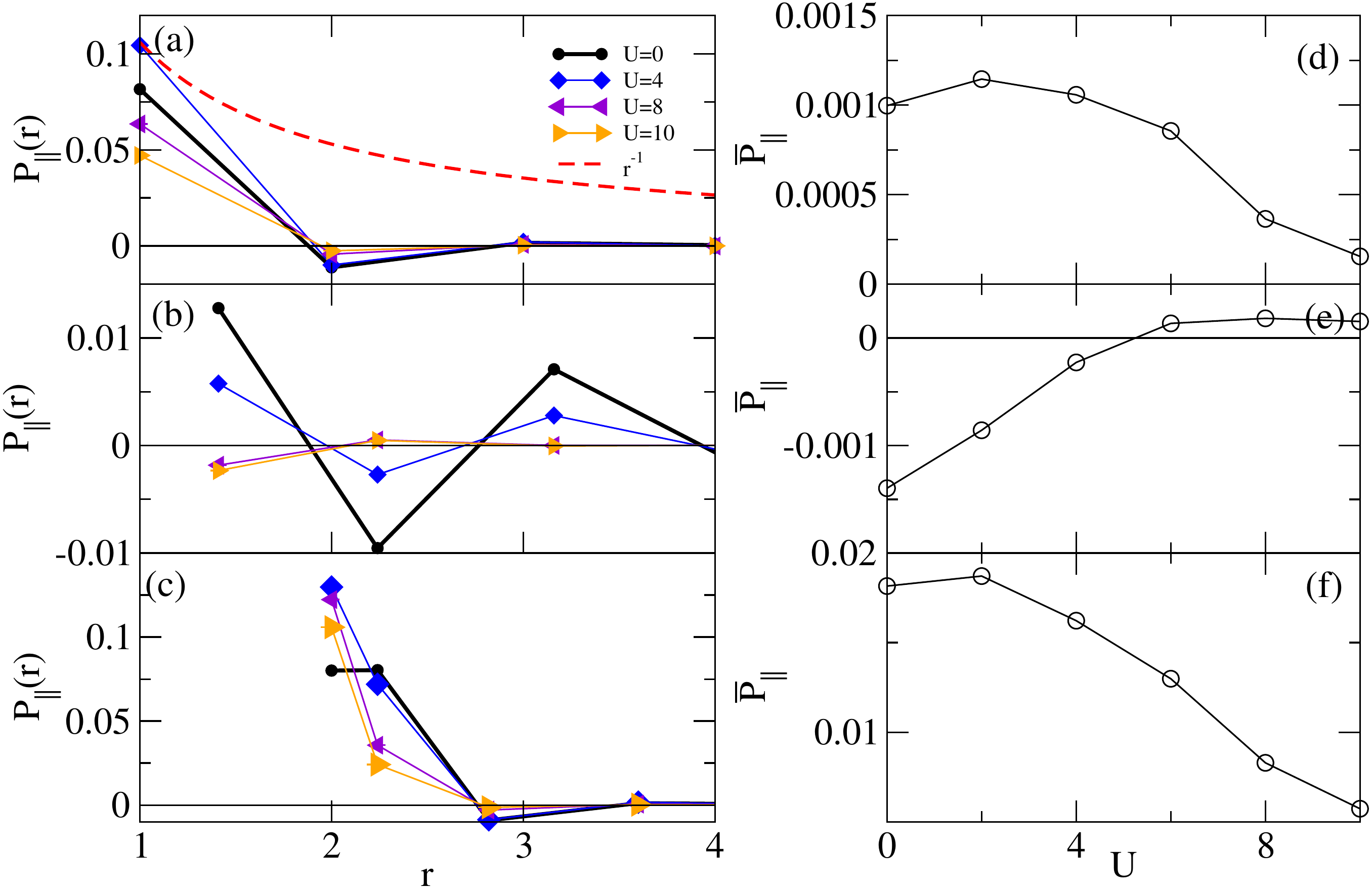}}
  \caption{Pair-pair correlation for $t^\prime=0.2$ for
    parallel-oriented n.n. pairs. (a) Pair-pair correlation
    $P_{||}(r)$ for n.n. $x$-axis pairs on chain 1 (b) $P_{||}(r)$ for
    n.n. $x$-axis pairs on chains 1 and 2 (c) $P_{||}(r)$ for
    n.n. $x$-axis pairs on chains 1 and 3. In (a)-(c), $r$ is the
    center-to-center pair distance.  In (a) we show the function
    $r^{-1}$ for comparison as the dashed curve.  In panels (d) we
    plot the $U$ dependence of the average long-range correlation
    $\bar{P_{||}}$ (see text) for the chain 1 -- chain 1 correlations
    in (a); panels (e) and (f) are similar for chain 1 -- chain 2 and
    chain 1 -- chain 3 correlations, respectively.}
\label{tp.2-x-x}  
\end{figure}

In Fig.~\ref{tp.2-x-x} we plot $P_{||}(r)$ as a function of $r$. In
Fig.~\ref{tp.2-x-x}(a) the indices $i$ and $j$ in Eq.~\ref{pr} are
taken on the same chain, in Fig.~\ref{tp.2-x-x}(b) they are on
neighboring chains, and in Fig.~\ref{tp.2-x-x}(c) they are on second
neighbor chains. In each panel $r$ is the center-to-center distance
between pairs in units of the dimer-dimer spacing. In
Fig.~\ref{tp.2-x-x}(a) we also plot the function $r^{-1}$.  While the
distances we have on the lattice are limited, and we cannot determine
if $P(r)$ decays with $r$ as a power law or as an exponential, we find
that on each chain $P(r)$ decays significantly faster than $r^{-1}$.

\begin{figure}
  \centerline{\includegraphics[width=4.75in]{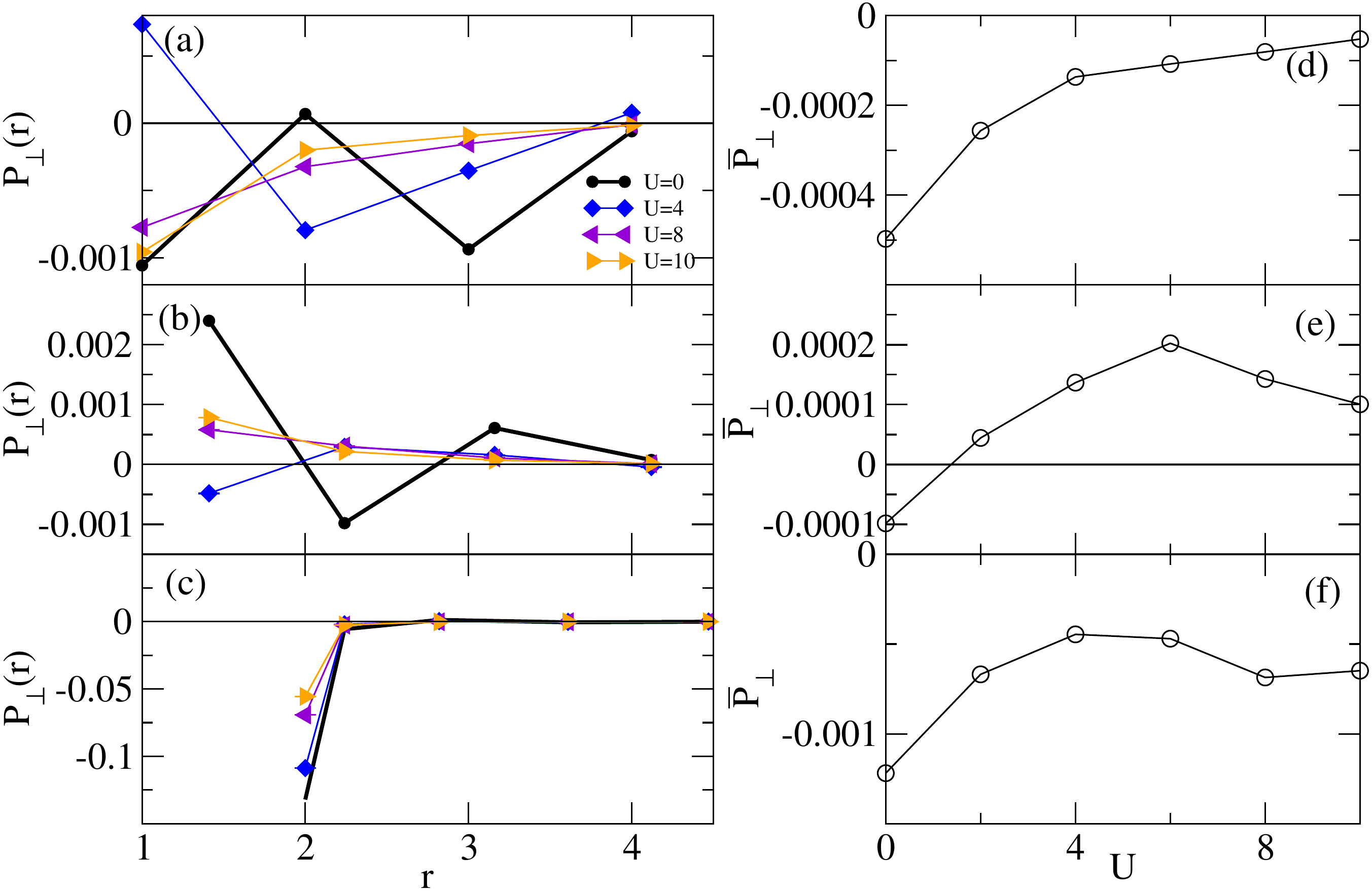}}
  \caption{Panels (a)-(f) are the same as in Fig.~\ref{tp.2-x-x}
    except that the pair-pair correlation $P_{\perp}(r)$ is for the
    perpendicular orientation of pairs. Note that in (d)-(f),
    $\bar{P}_{\perp}$ is negative, with decreasing magnitude as $U$ increases.
    $\bar{P}_{\perp}$ in (e) is strongly affected by shorter-range
  correlations which lead to the unusual $U$ dependence (see text). }
\label{tp.2-x-y}  
\end{figure}
As seen in Fig.~\ref{tp.2-x-x}(a) and (c), only at very short range
($r\leq 2$) we find that $P(r,U)$ is enhanced very slightly over its
value at $U=0$. This same effect is seen in calculations in the
monomer $\frac{1}{2}$-filled Hubbard model on an anisotropic triangular
lattice \cite{Clay08a,Dayal12a}. As noted in \cite{Aimi07a},
at $r=0$ the pair-pair correlation
can be written as a linear combination of charge-charge and spin-spin
correlations, which are enhanced by $U$ \cite{Aimi07a}.  To measure
the effect of correlations on pairing we define the average long-range
pair-pair correlation \cite{Gomes16a},
\begin{equation}
  \bar{P}=\frac{1}{N_p}\sum_{2<|\vec{r}|<r_{\rm max}}P(r).
  \label{pbar}
\end{equation}  
In Eq.~\ref{pbar} $N_p$ is the number of terms in the sum, and
the sum is over $r$ greater than 2 in dimer-dimer
units but less than a maximum distance $r_{\rm max}$. The upper cutoff
is necessary because of the open boundary conditions along $x$ in
our lattice and is chosen so that the maximum pair-pair distance
along the chain is 4 (one half of the lattice length of 8 dimers).
Fig.~\ref{tp.2-x-x}(d)-(f) show $\bar{P}_{||}$ as a function of $U$.
In each case the magnitude of  $\bar{P}_{||}$ decreases with $U$
apart from an insignificant increase at very small $U$. A large
decrease in $\bar{P}_{||}$ also takes place for $U\gtrapprox 8$
upon entering the AFM insulating phase. A similar decrease is
seen in calculations within the monomer $\frac{1}{2}$-filled band \cite{Clay08a,Dayal12a}.

Fig.~\ref{tp.2-x-y}(a)-(c) shows perpendicular correlations $P_\perp(r)$
as a function of $r$.
Like $P_{||}$, we find that $P_{\perp}$ along the chain direction decays
faster than $r^{-1}$.
These are negative in many cases
(Fig.~\ref{tp.2-x-y}(a) and (c)), which would be consistent with
a $d_{x^2-y^2}$ type pairing symmetry. Fig.~\ref{tp.2-x-y}(d) shows
a clear decrease in magnitude of $P_{\perp}(r)$ with increasing
$U$.  The $U$ dependence of Figs.~\ref{tp.2-x-y}(e)-(f) are harder
to interpret. We believe that the unusual $U$ dependence here
is due to shorter range correlations, for example as in Fig.~\ref{tp.2-x-y}(b)
where $P_\perp(r)$ changes discontinuously at small $U$
at $r\sim 2$ \cite{Clay19b}. If distances $r<3$ are excluded, we again find
a continuous decrease of $\bar{P}_\perp$ with increasing $U$.

\begin{figure}[H]
  \centerline{\includegraphics[width=4.75in]{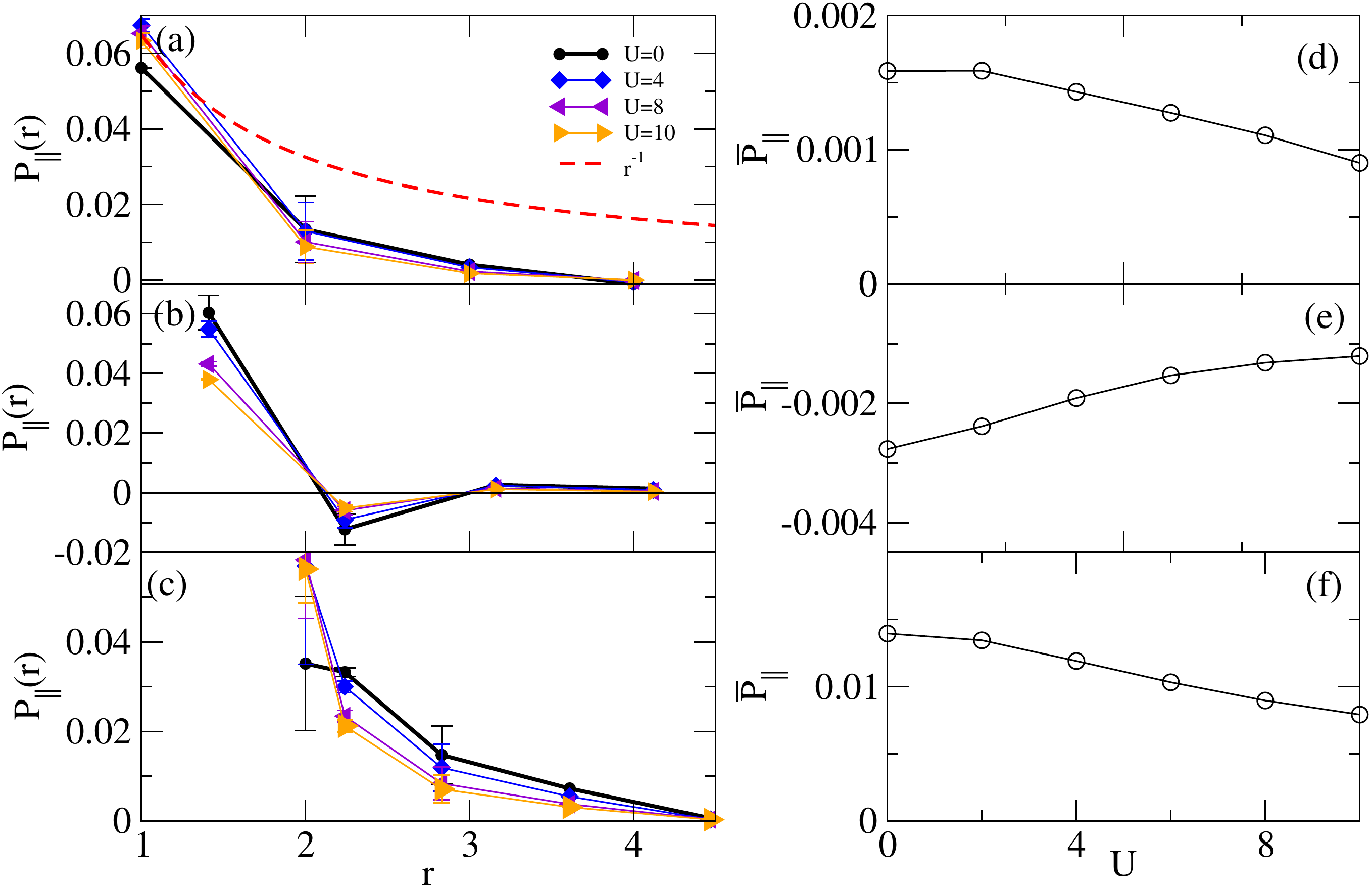}}
  \caption{Panels (a)-(f) are the same as Fig.~\ref{tp.2-x-x} but with $t^\prime=0.6$.}
\label{tp.6-x-x}  
\end{figure}
\begin{figure}[H]
  \centerline{\includegraphics[width=4.75in]{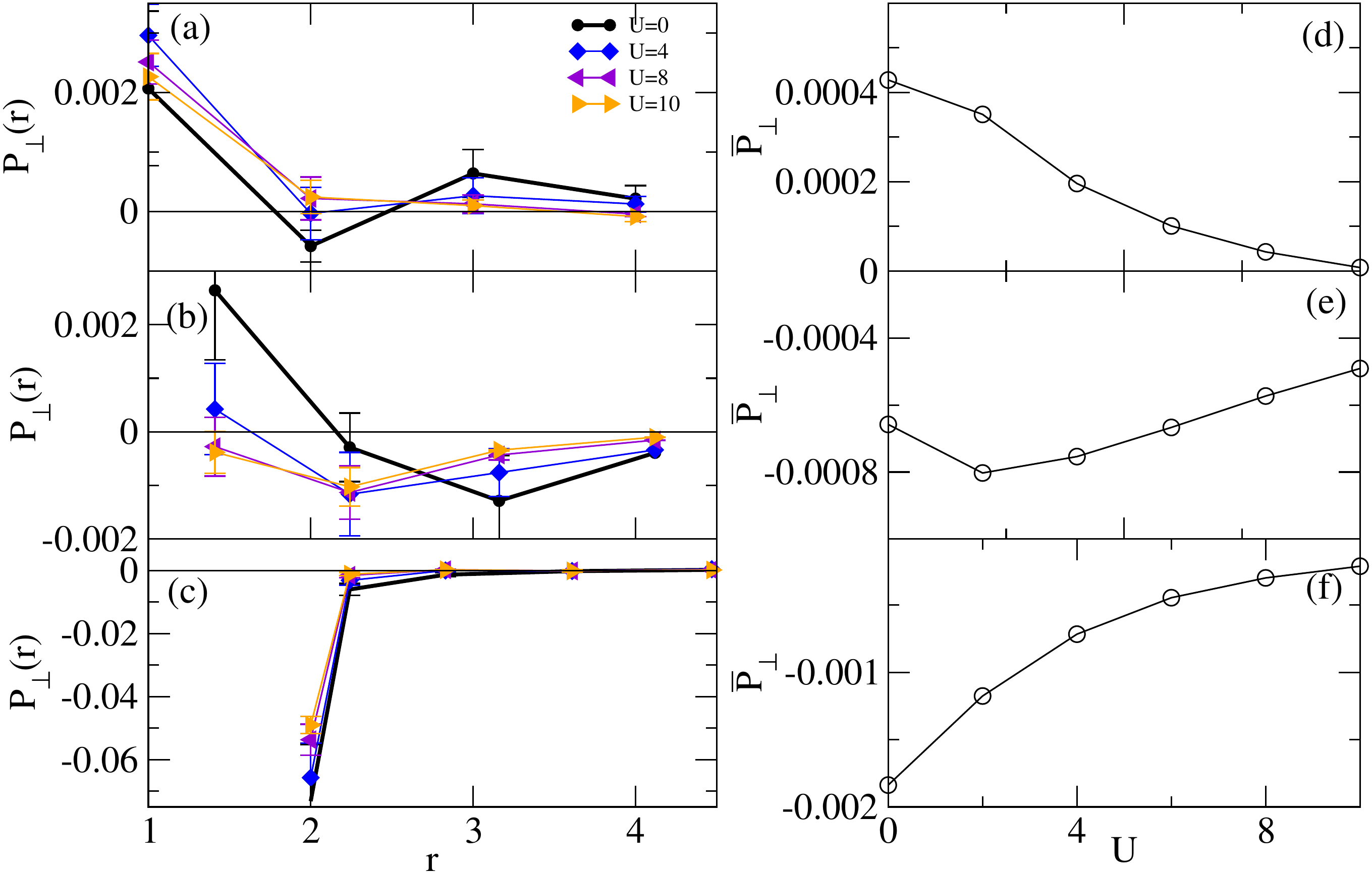}}
  \caption{Panels (a)-(f) are the same as Fig.~\ref{tp.2-x-y} but with $t^\prime=0.6$.}
\label{tp.6-x-y}  
\end{figure}

Figs.~\ref{tp.6-x-x} and Fig.~\ref{tp.6-x-y} show the pair-pair
correlations at $t^\prime=0.6$. Here AFM order is absent over the
whole range of $U$ we studied. We again find that $P_{||}(r)$
decreases faster than $r^{-1}$ (see Fig.~\ref{tp.6-x-x}(a)). As
at smaller $t^\prime$ we again find that some pair-pair correlations are
enhanced by $U$ at short range ($r\leq 2$). However, here both
$\bar{P}_{||}$ and $\bar{P}_\perp$ decrease continuously with $U$,
again with the exception of one point at shorter $r$ where there
is a discontinuous change at small $U$ (see Fig.\ref{tp.6-x-y}(b) and (e)).

Summarizing our data, while we cannot show the presence of long range
AFM order from our single lattice calculation, we do find strong AFM
correlations for weak frustration and moderately strong $U$.  We find
no evidence that within the effective $\frac{1}{2}$-filled picture
that the superconducting pair-pair correlations are enhanced over
their value for non-interacting electrons.

\section{Discussion}

The conclusions of our work are, (i) SC is absent within the model,
and (ii) the presence or absence of AFM order is not related to the
tendency towards SC.

In $\kappa$-(BEDT-TTF)$_2$X with X = Cu[N(CN)$_2$]Cl where AFM order
is found under ambient pressure, estimates for $t^\prime/t$ vary
between 0.44 and 0.53 \cite{Clay19a}.  Similarly, no AFM order is
found for X = Cu$_2$(CN)$_3$ where 0.80 $\lessapprox$ $t^\prime/t$
$\lessapprox$ 0.99 \cite{Clay19a}.  In our present results we find AFM
order at $t^\prime=0.2$ ($t^\prime/t=0.4$) but no AFM order at
$t^\prime/t=0.8$, which is roughly consistent with what is observed
experimentally.  SC is however found in {\it both} X = Cu[N(CN)$_2$]Cl
and X = Cu$_2$(CN)$_3$, under pressures of 0.03 and 0.6 GPa,
respectively.  Moreover, there is no obvious connection between the
superconducting $T_{\rm c}$ in $\kappa$-(BEDT-TTF)$_2$X and the value
of $t^\prime/t$ \cite{Clay19a}.  For example SC is also found for
$t^\prime/t>1$ ($t^\prime/t=1.3$ in X = CF$_3$SO$_3$
\cite{Ito16a}). This suggests that SC in $\kappa$-(BEDT-TTF)$_2$X is
not related to the proximity to an AFM or QSL phase. This is also
consistent with our present results, where we find little difference
in the pair-pair correlations between $t^\prime=0.2$ where AFM is
present and $t^\prime=0.6$ where AFM is absent.

Our results on the absence of SC in the effective $\frac{1}{2}$-filled
Hubbard dimer model is in complete agreement with our previous results
on the $\frac{1}{2}$-filled Hubbard monomer model
\cite{Clay08a,Dayal12a,Gomes13a}. Independent of whether one considers
the dimers of BEDT-TTF molecules as effective single sites with charge
occupancy +1 (previous calculations) or as dimers with monomer charge
occupancies of $\frac{1}{2}$ each (present calculations), the
superconducting pair correlations are found to be suppressed by
Hubbard $U$ continuously from $U=0$. This fundamental observation was
arrived at from exact diagonalization \cite{Clay08a,Gomes13a}, Path
Integral Renormalization Group \cite{Dayal12a} and now DMRG on
completely different lattices, thereby confirming that the result is
not an artifact of the choice of a specific lattice or a particular
computational technique. Taken together, these results very clearly
indicate that spin fluctuations are not driving organic SC. With
hindsight, this conclusion should not be surprising: as already
pointed out in the above, AFM occurs in only three
$\kappa$-(BEDT-TTF)$_2$X, X = Cu[N(CN)$_2$Cl, deuterated X =
  Cu[N(CN)$_2$Br and CF$_3$SO$_3$. Whether or not X = Cu$_2$(CN)$_3$
    and Ag$_2$(CN)$_3$ are true QSLs continue to be debated, and both
    claims of a CO phase below 6 K \cite{Kobayashi20a} as well its
    absence \cite{Sedlmeier12a} exist in the literature.  What is
    perhaps very relevant in this context that transition from CO to
    SC is far more common in CT solids, including in the $\beta$,
    $\beta^\prime$, $\beta^{\prime\prime}$ and $\theta$-families
    \cite{Clay19a}.  Even among the $\kappa$-phase materials, CO has
    been detected in X = Hg(SCN)$_2$Cl
    \cite{Drichko14a,Drichko20a,Gati18a} while in the related material
    X = Hg(SCN)$_2$Br there occurs a dipolar liquid phase
    \cite{Drichko18a} the formation of which is related to the
    mechanism of CO formation \cite{Hotta10a,Naka10a}.  While these
    latter materials are also strongly correlated, CO requires an an
    effective $\frac{1}{4}$-filled (or $\frac{3}{4}$-filled
    description of CT solids, within which intra-dimer charge dimer
    degrees of freedom are explicitly taken into account
    \cite{Clay19a,Li10a,Dayal11a,Seo00a,Kaneko17a}.  The charge-charge
    correlations in the CO phase in these systems are different from
    that in Wigner crystals. The CO phase is a paired-electron
    crystal, with nearest neighbor spin-singlet charge-rich molecules
    separated by pairs of charge-poor molecules. Pressure increases
    frustration, causing the spin singlets to be mobile.  Calculations
    on an anisotropic triangular lattice \cite{Gomes16a} as well as on
    $\kappa$-lattices \cite{DeSilva16a,Clay19b} have indeed found that
    superconducting pair correlations are enhanced by the Hubbard $U$
    uniquely at $\frac{1}{4}$-filling.  Additionally,
    electron-electron and electron-bond phonon interactions act
    co-operatively at this bandfilling to enhance superconducting pair
    correlations, even at they compete with one another at all other
    fillings \cite{Clay20a}.  These results are encouraging and
    suggest further research along this direction.

\funding{S.M. acknowledges support by the National Science Foundation grant NSF-CHE-1764152.}

\conflictsofinterest{The authors declare no conflicts of interest.}

\end{paracol}
\reftitle{References}

\externalbibliography{no}

\end{document}